\documentclass[conference]{IEEEtran}
\IEEEoverridecommandlockouts

\usepackage[utf8]{inputenc}
\usepackage[T1]{fontenc}
\usepackage{graphicx}
\usepackage{amsmath}
\usepackage{amsfonts}
\usepackage{amsthm}
\usepackage{amssymb}
\usepackage{balance}
\usepackage{cite}
\usepackage{units}

\usepackage{color}

\usepackage{ifthen}

\usepackage[unicode=true,
  hidelinks,
  pdfsubject={}, 
  pdfkeywords={},
  colorlinks=true,
  allcolors=blue, 
  citecolor=blue
]{hyperref}

\newboolean{blinded}
\setboolean{blinded}{false}

\newboolean{showcomments}
\setboolean{showcomments}{false}

\newcommand{\stb}[3][-]{\ifthenelse{\boolean{showcomments}}
  {\textcolor{red}{{\bf Story: (#2)}\\ #3\ \\ (Comments: #1)}}{}}

\ifthenelse{\boolean{showcomments}}
           { \newcommand{\mynote}[3]{
               \fbox{\bfseries\sffamily\scriptsize#1}
                    {\small$\blacktriangleright$\textsf{\emph{\color{#3}{#2}}}$\blacktriangleleft$}}}
           { \newcommand{\mynote}[3]{}}

\newcommand{\mv}[1]{\mynote{MV}{#1}{blue}}
\newcommand{\pjv}[1]{\mynote{PJV}{#1}{darkgreen}}
\newcommand{\ali}[1]{\mynote{Ali}{#1}{darkblue}}

\begin{document}

\title{The Path to Fault- and Intrusion-Resilient \\ Manycore Systems on a Chip}

\author{
  \ifthenelse{\boolean{blinded}}
             {Disrupt Paper No. 5073}{
  \IEEEauthorblockN{Ali Shoker \quad Paulo Esteves-Verissimo}
  \IEEEauthorblockA{
    RC3 Center, CEMSE Division, \\
    King Abdullah University of Science\\ 
    and Technology (KAUST)\\
    \url{ali.shoker@kaust.edu.sa}, \url{paulo.verissimo@kaust.edu.sa}}
    \and
  \IEEEauthorblockN{Marcus V\"olp}
  \IEEEauthorblockA{
    University of Luxembourg\\
    Interdisciplinary Center for Security, \\
    Reliability and Trust (SnT) - CritiX group\\
    \url{marcus.voelp@uni.lu}}
    \and
%      \IEEEauthorblockN{Paulo Esteves-Verissimo}
%  \IEEEauthorblockA{
%    RC3 Center, CEMSE Division, \\
%    King Abdullah University of Science\\ 
%    and Technology (KAUST)\\
%    \url{paulo.verissimo@kaust.edu.sa}}
  % \IEEEauthorblockN{Ali Shoker, Paulo Esteves-Verissimo}
  % \IEEEauthorblockA{
  %   King Abdullah University of Science and Technology (KAUST)\\
  %   Resilient Computing and Cybersecurity Center (RC3)\\
  %   \url{ali.shoker@kaust.edu.sa}, \url{paulo.verissimo@kaust.edu.sa}}
}}

\maketitle

\thispagestyle{plain}\pagestyle{plain}

\begin{abstract}
  \stb{147 out of 150 words}{write me}
The hardware computing landscape is changing. What used to be distributed systems can now be found on a chip with highly configurable, diverse, specialized and general purpose units. Such Systems-on-a-Chip (SoC) are used to control today's cyber-physical systems, being the building blocks of critical infrastructures. They are deployed in harsh environments and are connected to the cyberspace, which makes them exposed to both accidental faults and targeted cyberattacks. This is in addition to the changing fault landscape that continued technology scaling, emerging devices and novel application scenarios will bring. In this paper, we discuss how the very features---distributed, parallelized, reconfigurable, heterogeneous---that cause many of the imminent and emerging security and resilience challenges, also open avenues for their cure though SoC replication, diversity, rejuvenation, adaptation, and hybridization. We show how to leverage these techniques at different levels across the entire SoC hardware/software stack, calling for more research on the topic.
% In this paper, we highlight these challenges, argue why adaptation is key to their solution and why adaptation is necessary even at the lowest levels and must be supported throughout the hardware/software supply chain.
\end{abstract}

\pjv{What would be the differences to the Midir paper that make this a disrupt?}
\mv{I would consider Midir only the basic mechanism (of course selling a bit of it). What we need is core rejuvenation SW + HW and for that diversity and morph around persistent faults (othw. we loose cores over time). Also there are a lot of open questions at the OS side, like how to benefit from privilege reversion, what mechanism exactly (e.g., iBFT = Midir, but at different levels).}


%\section{Challenges}
%\label{}

\section{Opportunities for hardware resilience}

\ali{need to expand on the aging issues, .. trojans, and attacks in hardware}

Hardware chips continue to be the core building blocks of computing devices due to their inherent immutability and speed, required in modern digital and mission-critical systems like Cyber-Physical Systems, Healthcare, Fintech, Automotive, and Space. This hardware can implement an entire monolithic system or even be used as proof-of-trust anchors. Contrary to the common belief, hardware is prone to unintentional (benign) and intentional/malicious (intrusion or Byzantine~\cite{pbft:1999}) faults. The former can be caused by the fabrication (e.g., Silicon) material prone to dust, aging, and overheating, or by design/implementation glitches~\cite{merlino2004dusty, celaya2010accelerated}. Malicious faults manifest in many forms, prior- or post-fabrication, where stealthy logic, \textit{backdoors}, \textit{trojans}, \textit{kill switches}, and post-fab fabric editing are possible~\cite{attack-adee2008hunt,attack-imeson2016non,attack-king2008designing,attack-yang2016a2}. In line with this, the trends of building complex hardware out of smaller \textit{commercial-off-the-shelf} (COTS) components and introducing programmable/reconfigurable hardware, e.g., FPGA~\cite{FPGA-2008fpga,Xilinx2019}, are closing the gap with software systems: hardware systems are no longer rigid, immutable, and fixed creatures. This raises both new challenges and opportunities, which call to revisit the way resilient and secure hardware systems are built.

The notable demand on hardware due to the automation and digitalization of services in many sectors raised new challenges in the hardware fabrication industry, where vendors need to maintain delivery on time and reduce production costs. This resulted in a \textit{divide-and-conquer}~\cite{divide-conquer-1980} production style: a system is split into smaller and cheaper building blocks, i.e., components. Components are developed in parallel to reduce the production cycle time. Each block is likely developed by a dedicated specialized vendor, i.e., generating COTS~\cite{COTS-1998opportunities}. This means that the synthesising entity of these COTS can focus on the technology it masters, rather than distributing its efforts on multiple fronts. Despite this, these cheap components are becoming more prone to failures and attacks~\cite{COTS-issues-2006}, which can lead to drastic impacts on critical sectors like Cyber-Physical Systems, health smart systems, mission-critical space systems, etc. Our experience in software systems shows that building resilient systems composed of small and cheap components can be more resilient than a single complex monolithic system, that is usually very expensive.

\begin{figure}[t]
\centering
\includegraphics[width=0.8\columnwidth]{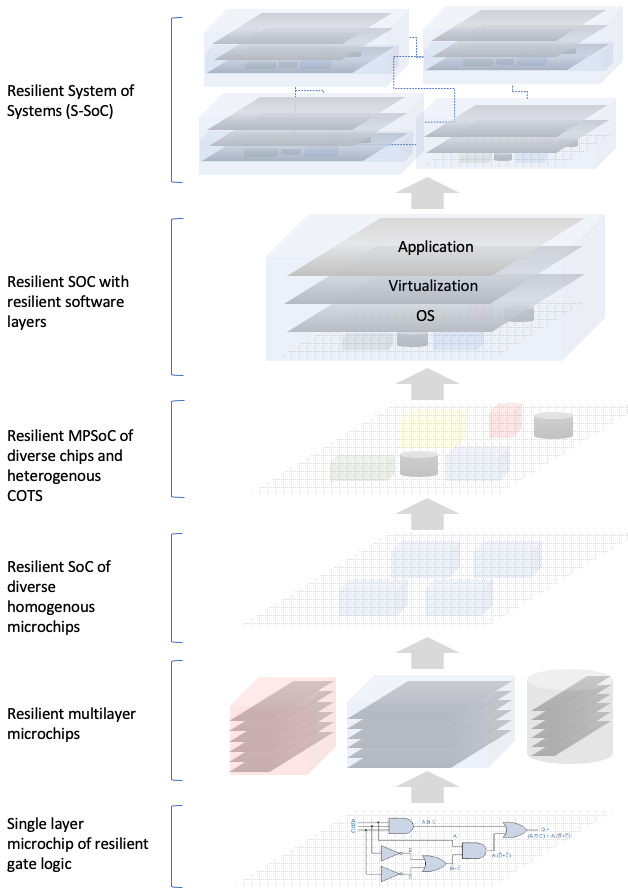}
\caption{ Resilience forms at the different (networked) hardware layers of Multicore Systems on Chip.}
\label{fig:res-chips}
\end{figure}

There are ample opportunities for hardware resilience leveraging the above advancements. To demonstrate this, we showcase in Fig.~\ref{fig:res-chips} different levels of the chip development process, from low-level fine-grained gate logic blocks up to multicore systems-on-chip (SoC). Literary works reveal some selected resiliency techniques on most of these layers for constructing resilient clock networks, replicated power domains, and lock-step coupling of cores~\cite{gate-redundancy-2009,TMR-1962,TMR-automated-2018,TMR-FPGA-comparison2007,gate-res-2016,evolution-circuit-evaluation-1999}, which is a good starting point. We, however, advocate for more systematic and comprehensive resiliency, probably leveraging hardware \textit{hybrids} to simplify the designs. This holistic view helps optimising SoC designs by suggesting the right level of resiliency at each stage to reduce the redundant complexity and cost.

In a nutshell, the lowest level, in Fig.~\ref{fig:res-chips}, is building a single layer microchip that constitutes a simple logical circuit of gates. Different gates are known to have different resiliency levels~\cite{gate-redundancy-2009,gate-res-2016}. Recently, SiNW transistors are used to bridge \textit{Source} to \textit{Drain} with multiple \textit{nanowires} to compensate manufacturing defects and aging~\cite{sinw_array}. While a typical design process mainly considers the \textit{space}, \textit{energy}, and \textit{time} metrics in the design, making these circuits more resilient would mean trading these metrics for resiliency, e.g., using backup gates, replicated parallel gates, or diverse gates~\cite{gate-res-2016,evolution-circuit-evaluation-1999}. 
%
%This 
On the other hand, single-layered circuits can today be synthesized in a \textit{3D fabric}\cite{3D-book-2017}. Layers typically have different complementary functionalities. However, they can also have layers of identical functionality from different vendors, which is useful to improve diversity in fault masking scenarios (discussed later). It is also helpful to synthesize a monolithic chip from multi-vendor layers to avoid vendor lock-in or potential aging issues, backdoors, and \textit{kill switches}~\cite{attack-adee2008hunt,merlino2004dusty,celaya2010accelerated}---so called \textit{Distribution attack} on the supply chain.

At a higher level, always depicted in Fig.~\ref{fig:res-chips}, these 3D microchips can be assembled to build a system-on-chip fabric~\cite{soc2001}. Again, components of identical functionalities can be used to build fault and intrusion masking SoC fabric. This can be enriched with heterogeneous diverse microchips at a higher level, thus building resilient Multicore Systems on Chip (MPSoC)~\cite{mpsoc-2008,Xilinx2019}.
At the higher layers, where a software stack complements the functionality of the system to form a more \textit{programmable} flexible hardware (discussed next), one can take advantage of a remarkable body of research and practice to build resilient \textit{soft-custom} logic~\cite{diverse-virtualization-2008,BFT-SDN-2016,intrusion-virtualization2010}. This can be done by exploiting virtualization techniques to provide software-level containment and replication.
More complex systems can be built through networked \textit{systems of systems on chip}. First instances of \textit{networked} SoC systems are already emerging in the automotive, aeronautics, and CPS domain.

Across this spectrum, we foresee a need and opportunities to revisit how resilient hardware is built:
%, leveraging the following:
\begin{itemize}
    \item building complex systems of systems and MPSoCs out of smaller COTS;
    \item taking advantage of the programability and elasticity of modern hardware, e.g., FPGA, GPGPU, to replicate, diversity and adapt; and
    \item simplifying the design of secure robust systems using smaller hardware \textit{hybrids}---easy to design and verify, as resilient anchors.
\end{itemize}

\stb{1/2 pg}{
  Challenge 1: fault containment
  \begin{itemize}
  \item SPOF energy supply: different domains
  \item SPOF clock network: GALS + Steininger's resilient clock work
  \item crosstalk
  \item heat as cross-core fault injection; side- / covert channel
  \item ...
  \item => must construct systems such that at some level only
    accidental faults can happen; then error detection / correction to
    constrain how faults propagate; the lower this level, the better,
    but this also implies faults might propagate into legitimately
    accessible objects
  \end{itemize}
}

\pjv{shouldn't w talk abt fault independence as well?}
\mv{Yes, at least in terms of requirements and solutions that exist (Steininger's clock algorithm, heterogeneous cores, ultimately leading to reconfiguration and organic computing as primitives)}

\stb{1/2 pg}{
  Challenge 2: need for operational resilience
  \begin{itemize}
  \item SOTA: static partitioning, but modern apps require change
  \item change to adapt to threats, to adapt to
    application changes, to safe power
  \item impossibility to support change without trusted components:
    change requires changing access to resources, but the entity
    performing that change and the entity enforcing this change may
    fail as well. In particular a failure in any of the two may grant
    access to a resource that shouldn't be accessed and faults may
    propagate into that resource => impossibility
  \item it suffices for enforcement to be trusted, provided the
    decision what to enforce is made consensually <= Midir \cite{gouveia2022behind}
  \item change is also required to recover from faults
  \item relocate, rejuvenate
  \end{itemize}
}
\section{Programability, Elasticity, Plasticity}
\label{sec:reprogram}

The genuine 
%hard-
immutability properties of hardware components and elements, make them ideal for security hardening and containment, i.e., by making the hard-implemented logic tamper-resistant against both benign and intrusion faults.
Despite these facts, there is a continuous wave of relaxing these “rigid” hardware designs through introducing programmable (including reconfigurable and adaptable) fabric~\cite{FPGA-2008fpga,gpgpu-CUDA-2012gpgpu}. The main reason is to improve hardware flexibility and compatibility, i.e., making them application-agnostic, and to facilitate the daunting design verification process prior to fabrication, hence cutting off fabrication costs thereof. For this, programmable hardware is considered a tradeoff between software logic---fully flexible, slow, and mutable---and hard logic---fully rigid, fast, and immutable. 
%Of particular interest, 
We believe that there are promising opportunities to boost the resilience of the programmable platforms against faults and intrusions, although immutability is slightly reduced. To explain these benefits, we consider two classes of programmable hardware:

\textbf{Soft Custom Logic Fabric (SCLF)} : these are commonly known as software-defined devices like PLC, ECU, and SDN devices~\cite{vPLC-2021,plc-sec2021security,ecu-2021survey,BFT-SDN-2016}. This hardware is mostly domain-specialized, where computing is done using general-purpose micro-controller or microprocessors, often managed by a full software stack: hypervisors, RTOS/OS, drivers, libraries, and applications. Consequently, these devices exhibit high programability features, analogous to IT computing, although they have specialized roles and use domain specific peripherals, e.g., sensors, actuators, and interfaces.

% 

% this level has a significant software role that complements the functionalities of the hardware, e.g., Software-defined network or Software-defined Electronic Control Units GPGPU. Hardware is used for general purpose computing and providing secure root of trust: random number generation (TRNG), secure boot mechanisms, secure update, secure debug, cryptographic acceleration, tamper resistance and protection of secrets, tamper detection, on-the-fly memory encryption, process/functions isolation, and run-time integrity protection. Software here can cover the entire IT software stack from driver, though OS, and a;; the way to applications. This makes these classical devices highly programmable, configurable and adaptive to different applications and contexts.

\textbf{Hard Custom Logic Fabric (HCLF)} : these are hardware chip fabrics, e.g., FPGA~\cite{FPGA-2008fpga} and GPGPU~\cite{gpgpu-CUDA-2012gpgpu}, composed of arrays of logical components, e.g., \textit{gates} and \textit{multiplexers}, that are not “hard etched”, i.e., can be reprogrammed as needed. The programming logic in this case is almost entirely implemented in hardware, without the need for a software stack at runtime. Fabric is reprogrammed through soft \textit{IP Cores}~\cite{xilinx-IP-core,Intel-IP-core}
%\footnote{https://www.intel.com/content/www/us/en/products/details/fpga/intellectual-property.html OR https://www.xilinx.com/products/intellectual-property.html} 
(HDL code~\cite{hdl-1998jhdl}) or through components (\textit{softcores} or \textit{blocks}) synthesized on the chip as needed. This programability feature is a very interesting tradeoff that retains the speed and security of \textit{Application-Specific Integrated Circuit} (ASIC) chips, while giving the flexibility to support diverse applications and update implementations without the need for costly and slow fabrication.%\\

Although programability, in both classes, opens the door for tampering with the system, and thus injecting surveillance circuits, intrusions and backdoors~\cite{attack-adee2008hunt,celaya2010accelerated} after fabrication (though slightly compared with software systems), there is a huge opportunity to leverage this programability to improve the resilience of these systems through four main ingredients: replication, diversity, rejuvenation, and adaptation.

\subsection{Replication}
Replication is often useful to build resilience against Benign or Byzantine faults. \textit{Passive replication}~\cite{passive-backup-1993primary,semi-passive-rep-1998} allows a failing system to failover into a backup replica. This is a cheap solution that typically requires one passive backup replica. However, recovery is slow, requires reliable detection and is not seemless to the user, even if implemented entirely at transistor level. 
For example, Razor~\cite{1253179} integrates detection capabilities, originally for timing faults in sequential logic, but also for power instability~\cite{1610623} and side channels~\cite{6487728}, and re-injects stored state into the pipeline for re-execution. Albeit functionally transparent, users may observe timing differences and anomalies caused by them. 
\textit{Active replication} masks faults through building a \textit{deterministic replicated state machine}~\cite{smr:1993}, composed of replicas of identical functionality, which execute an agreement protocol, e.g. \textit{Paxos}~\cite{lamport2001paxos} or \textit{PBFT}~\cite{pbft:1999}. The number of required replicas is typically $2f+1$/$3f+1$ in order to tolerate $f$ faults. Interestingly, several works make use of hardware hybrids as root-of-trust to simplify these protocols to build resilient broadcast and agreement abstractions for embedded real-time systems~\cite{efficient-hybrid-BFT-2016,minBFT-2011efficient,pistis:2021} (requiring only $2f+1$ replicas to tolerate $f$ Byzantine ones). 

Replication in SCLF is analogous to software replication at the software layer. While some literary works have studied this in some settings~\cite{BFT-SDN-2016,diverse-virtualization-2008,intrusion-virtualization2010,IRS-vehicle-2022intrusion}, there are research opportunities in other real-time applications like software-defined vehicles, UXVs, Smart Grid, etc.
On the other hand, replication in HCLF is today easier than ever. Using an FPGA, it is possible to spawn replicas as soft cores or logical blocks, using off-the-shelf soft IPs. This is a nice hardware feature that gives the flexibility to create hard-replicas quickly and on-demand, using only one fabric, in a similar way to creating virtual machines or containers at software level.  

\subsection{Diversity}
Resiliency through active replication is, however, only guaranteed as long as the replicas fail independently~\cite{pbft:1999,smr:1993}. The second ingredient, diversity, helps building replicas of the same functionality but with different implementations. The aim is to avoid common-mode benign failures and intrusions.

Since programability in both classes, SCLF and HCLF, open new avenues for multi-vendor implementations and COTS, the likelihood of diversity is higher than the case of monolithic hardware that require deep technology capabilities. An interesting trend that would benefit this model greatly is more standardization for architectures and APIs. For instance, the introduction of the \textit{AutoSAR}~\cite{autosar-2023} standard has greatly enriched the automotive market with multi-vendor implementations of the entire software and hardware stack, which act as a blockboxes of identical functionalities. 
%Function Block diagrams are generic ready-to-use software COTS for PLCs. 
\textit{CUDA}~\cite{gpgpu-CUDA-2012gpgpu} and \textit{OpenGL}~\cite{opengl} provide standard APIs to implement accelerated parallel computing logic on a GPGPU using COTS implementations.
Open source hardware platforms like RISC-V~\cite{risc-v} also standardize the architectures provided by different vendors, and enrich the market with diverse architectures.

Interestingly, FPGAs allow for hardware diversity through modifying the hard-logic through using different implementations or specifications for the softcore/block IP, possibly from different vendors, which is then used to spawn computing cores. It would be interesting to study the case where IP compilers can generate diverse versions of identical softcores to be used on the fly. First approaches towards such a generation of morphable softcores has been investigated in the context of organic computing~\cite{5479525}.

\subsection{Rejuvenation}
Rejuvenation is the third complementary ingredient to replication and diversity. These latter techniques can only maintain resilience as long as the assumed number of failing replicas $f$ is fixed. This assumption is unfortunately hard due to benign faults and malicious behaviours. The first is related to aging, which manifest in software~\cite{soft-rejuvination-1995software} as memory leakage, failure to release resources and locks, failure to garbage collect, data corruption, etc. Surprisingly, aging occurs also in hardware, due to the deterioration of  hardware material under overuse and overheating, etc. The second reason is recently getting more attention with the increasing attempts of \textit{Advanced Persistent Attacks} (APT)---where a big deal of time and effort is usually put to identify vulnerabilities and exploit them. While this might be clear at the software level, there are continuous concerns about hardware backdoors and timed Trojans. Indeed, this is behind the recent agendas of acquiring chip sovereignty or split manufacturing in many countries~\cite{split-2020secure,soverign-2022digital}. 

SCLF reprogramability can greatly benefit from the huge body of research on software rejuvenation, that is proven to mitigate failures. This would even be more effective when rejuvenation is simultaneous with diversity, which allows the rejuvenation to a different implementation with identical functionally, in consequence, reducing the success rate of APTs. Using FPGAs, rejuvenation can also happen at hardware level in HCLF~\cite{Soc-rejuvination-2023system}. An FPGA allows restarting or spawning new soft cores and logical blocks at runtime---avoiding slow device restarts. In fact, one can partially rejuvenate some soft cores while others continue to run. FPGAs allow for even smarter techniques, e.g., to rejuvenate to diverse softcore variants that are loaded in different FPGA spatial locations, which can avoid potential backdoors in the FPGA grid fabric.

\subsection{Adaptation}
Yet, another way to withstand a varying number of faults $f$ is to adapt the resilient system accordingly. Among the adaptation forms are scaling out/in the system when $f$ may change, e.g., upon experiencing more threats, or switching to a backup protocol that is more adequate to the current conditions~\cite{adaptive-ft-smr2021,adapt-2015making,diverse-virtualization-2008,adapt-SDN-2018morph} (considering safety, liveness, performance, etc.). This would require research on the aforementioned adaptation mechanisms and, importantly, on severity detectors that can trigger adaptation actions once needed. As we discussed above, both SCLF (e.g., virtualization) and HCLF (e.g., FPGAs) provide tempo-spatial elasticity, which allows changing the number of replicas and their locations on the fabric as needed. It will be interesting to study these research questions from scratch or validating the feasibility of existing ones (developed in the software realm).

%In addition to the rejuvenation of replicas, FPGAs can be exploited to adaptively scale out the system to tolerate more than f faults when needed. This is practically considers the fabric as an elastic infrastructure where a varying number of hard replicas can be run on the FPGA when threat severity change over time.

\subsection{Resilient Reconfiguration}
\mv{or section}
It should be evident that reconfiguration must be resilient to faults and attacks, irrespective of the kind of adjustment performed (i.e., diverse rejuvenation, relocation, or adaptation). This holds for both reconfiguration of an FPGA grid fabric as well as multi-chip FPGAs---where the individual FPGA chiplets are the unit of reconfiguration. We shall focus here exclusively on internal, partial and dynamic reconfiguration, % and use these terms as well in their natural extension to multi-chip FPGAs where the individual FPGA chiplets are the unit of reconfiguration. 
%\textcolor{red}{Explain why it should be like that, e.g.: >>}
since the reliance on external complex and non-configurable modules (e.g., CPUs) would induce a weak spot in the system, which could contaminate its resilience or introduce downtimes. Nevertheless, dependencies on external \textit{hybrids} that are simple, and thus easy to verify, are allowed if they simplify the design. 
%since these   complex logic non-reconfigurable or introduce downtimes, which might not be tolerable (in particular in real-time systems).
%\textcolor{red}{<<}
Internal, partial and dynamic mean respectively that reconfiguration (i) is driven from within the FPGA, e.g., by an HCLF or softcore defining the configuration bitstream to be loaded into a reconfigurable region (or frame) through interfaces like internal configuration access ports, (ii) it is bound to the reconfigured area and elements therein, and (iii) it happens while other parts of the FPGA continue to execute.

%Although the final act of configuring is rather trivial, 
Optimizing the mapping of blocks to the FPGA grid fabric and integrating the configured block with the remaining blocks remain sufficiently complex tasks to be executed by a software-level operating-system kernel. Disabling and enabling configured circuits and frames constitute the critical operations, which leaves writing the configuration memory and validating that a correct bitstream is written as tasks that can be executed by the responsible kernel or possibly even kernel replicas. Provided sufficient access controls are in place at the internal configuration access ports, the actual configuration of a frame can even be delegated to its current user. However, as shown in Gouveia et al.~\cite{gouveia2022behind}, privilege change must remain a trusted operation executed \textit{consensually} and enforced by a trusted-trustworthy component. This leads to the more general question of architectural hybridization, which we address next.

\mv{
\begin{itemize}
    \item discuss mapping + integration with functionality in other reconfigurable regions
    \item complexity of mapping
    \item what are the critical operations: enabling of the reconfigurable region and the circuit installed therein + who can write which region => consensual reconfiguration by writing and validating configuration memory prior to enabling
    + consensual enabling interface
    \item actual reconfiguration may be delegated
    \item if so access controls to limit who can write a reconfigurable region's configuration memory
\end{itemize}}

\section{Architectural Hybridization}

\mv{split off reconfiguration access control from more general hybridization that Ali suggests and write about the latter}

\ali{more generic? In hybridization one would benefit from small easy-verifiable hybrids that are trusted-trustworthy. In particular, they can be used as anchors for trust or performance, which makes larger systems efficient, easy to design and verify. These could be unique counters (USIC), registers, secure elements, trusted sync networks, TPM, TEEs? The section now talks a lot about privileges and Midir, too specific. Let's widen our imagination ;)}

Differentiating how the individual hard- and software components of an MPSoC architecture can fail, architectural \textit{hybridization} aims at benefiting from small easy-to-verify and therefore more trustworthy components, called \emph{hybrids}. The goal is to enable, simplify or improve the performance of the overall system, by serving as trust anchors for these properties. These could be components (registers, memory, trusted execution environments or networks) such as USIG, A2M, TrInc, SGX and others, used in hybrid BFT-SMR protocols~\cite{minBFT-2011efficient,behl2017-hybrid,A2M-chun2007attested-hybrid,gupta2022dissecting-hybrid,kapitza2012cheapbft-hybrid, teechain-hybrid,jacobtrinc-hybrid,aguilera2023ubft-hybrid,7307998,1353018,10.1145/3492321.3519568}.
%, but also trusted synchronous network channels, trusted platform modules or entire trusted-execution environments, preferably in the style of IBM's secure co-processor for the S370 architecture.

Realizing hybridization poses a challenge dual to the question whether SCLF or HCLF leads to more reliable systems-on-a-chip. For software-only hybrids, we used to equate simplicity (measured for example in lines-of-code required to realize a certain functionality) with a low likelihood of failure and ease of verification. However, at hardware level, this equation is not as obvious, even if we consider lines-of-VHDL or another hardware description language. 

We illustrate this using the USIG from the MinBFT protocol by Veronese et al.~\cite{minBFT-2011efficient} as example. USIG is essentially a sequential circuit, which is driven by the counter register and a few additional registers, which provide as constants the secret key for the HMAC and the ID of the replica. The lowest complexity version of such a circuit will use normal registers. But then any bitflip in the counter will have catastrophic effects on the consensus problem at hand since it is reflected unchanged in the computed HMAC and USIG output. \textit{ECC-registers} on the other hand add extra bits and the logic required for correction, which both increase the complexity of the circuit at the benefit of tolerating a certain number of bitflips. We also see the converse effect when the required complexity of producing a special purpose circuit for a given functionality exceeds the complexity of a simple core that is able to fetch, decode and execute software. Once the inherent complexity of such a functionality exceeds this bound, software implementations become preferable and hybridization amounts to providing such an isolated core.

The objective of hardware-level hybridization is therefore to remain in this middle-ground. Hardware hybrids, protected by ECC and other accidental- and malicious-fault countermeasures, provide the desired functionality. This can then be extended into the realm of software hybrids that are possibly executed in a replicated manner and that vote to perform critical operations~\cite{gouveia2022behind}.

\section{Conclusions and Call to Action}
\label{sec:conc}

We emphasized that hardware architectures, and in particular multi- and manycore systems-on-a-chip are not the robust, dependable and reliable computing units we would like to have. We have subsequently started to replicate entire systems, which has ultimately lead to the huge body of knowledge on implementing resilient distributed systems. However, as we have seen, the continuing miniaturization and integration of processing elements into a single MPSoC, makes full system resilience increasingly costly, in particular when a single system already provides all the processing power that future critical applications need. We have shown how reconfiguration, rejuvenation and adaptation already allow the hardware to repair itself, to recover from faults and retain the resources classical resilience mechanisms need, when applied entirely on chip. 
%\textcolor{red}{We should maybe end by clearly stating our vision: e.g., >>}
Hybridization rooted in exactly the right-complexity circuits and applied to construct incrementally more complex dependable systems will produce the next generation flexible, morphable and highly trustable systems mission-critical systems will need. 
%\textcolor{red}{<<}
%
We therefore appeal for more research to study the resilience of hardware-based systems, systems of systems, and MPSoCs at different layers and cutting vertically across layers, probably through validating the techniques developed in the software \textit{Systems} and \textit{Dependability} areas.
\newpage

\bibliographystyle{IEEEtran}
\balance
\bibliography{resilient_socs}

\begin{thebibliography}{10}
\providecommand{\url}[1]{#1}
\csname url@samestyle\endcsname
\providecommand{\newblock}{\relax}
\providecommand{\bibinfo}[2]{#2}
\providecommand{\BIBentrySTDinterwordspacing}{\spaceskip=0pt\relax}
\providecommand{\BIBentryALTinterwordstretchfactor}{4}
\providecommand{\BIBentryALTinterwordspacing}{\spaceskip=\fontdimen2\font plus
\BIBentryALTinterwordstretchfactor\fontdimen3\font minus
  \fontdimen4\font\relax}
\providecommand{\BIBforeignlanguage}[2]{{%
\expandafter\ifx\csname l@#1\endcsname\relax
\typeout{** WARNING: IEEEtran.bst: No hyphenation pattern has been}%
\typeout{** loaded for the language `#1'. Using the pattern for}%
\typeout{** the default language instead.}%
\else
\language=\csname l@#1\endcsname
\fi
#2}}
\providecommand{\BIBdecl}{\relax}
\BIBdecl

\bibitem{pbft:1999}
M.~Castro, B.~Liskov \emph{et~al.}, ``Practical byzantine fault tolerance,'' in
  \emph{OsDI}, vol.~99, no. 1999, 1999, pp. 173--186.

\bibitem{merlino2004dusty}
R.~L. Merlino and J.~A. Goree, ``Dusty plasmas in the laboratory, industry, and
  space,'' \emph{PHYSICS TODAY.}, vol.~57, no.~7, pp. 32--39, 2004.

\bibitem{celaya2010accelerated}
J.~R. Celaya, P.~Wysocki, V.~Vashchenko, S.~Saha, and K.~Goebel, ``Accelerated
  aging system for prognostics of power semiconductor devices,'' in \emph{2010
  Ieee Autotestcon}.\hskip 1em plus 0.5em minus 0.4em\relax IEEE, 2010, pp.
  1--6.

\bibitem{attack-adee2008hunt}
S.~Adee, ``The hunt for the kill switch,'' \emph{IEEE Spectrum}, vol.~45,
  no.~5, pp. 34--39, 2008.

\bibitem{attack-imeson2016non}
F.~Imeson, S.~Nejati, S.~Garg, and M.~Tripunitara, ``$\{$Non-Deterministic$\}$
  timers for hardware trojan activation (or how a little randomness can go the
  wrong way),'' in \emph{10th USENIX Workshop on Offensive Technologies (WOOT
  16)}, 2016.

\bibitem{attack-king2008designing}
S.~T. King, J.~Tucek, A.~Cozzie, C.~Grier, W.~Jiang, and Y.~Zhou, ``Designing
  and implementing malicious hardware.'' \emph{Leet}, vol.~8, pp. 1--8, 2008.

\bibitem{attack-yang2016a2}
K.~Yang, M.~Hicks, Q.~Dong, T.~Austin, and D.~Sylvester, ``A2: Analog malicious
  hardware,'' in \emph{2016 IEEE symposium on security and privacy (SP)}.\hskip
  1em plus 0.5em minus 0.4em\relax IEEE, 2016, pp. 18--37.

\bibitem{FPGA-2008fpga}
I.~Kuon, R.~Tessier, J.~Rose \emph{et~al.}, ``Fpga architecture: Survey and
  challenges,'' \emph{Foundations and Trends{\textregistered} in Electronic
  Design Automation}, vol.~2, no.~2, pp. 135--253, 2008.

\bibitem{Xilinx2019}
Xilinx2019, ``Ug1085: Zynq ultrascale+ device technical reference manual,''
  \emph{Xilinx}, 2019.

\bibitem{divide-conquer-1980}
J.~L. Bentley, ``Multidimensional divide-and-conquer,'' \emph{Communications of
  the ACM}, vol.~23, no.~4, pp. 214--229, 1980.

\bibitem{COTS-1998opportunities}
L.~Brownsword and T.~Oberndorf, ``The opportunities and complexities of
  applying commercial-off-the-shelf components.''

\bibitem{COTS-issues-2006}
D.~Doan, ``Commercial off the shelf (cots) security issues and approaches,''
  NAVAL POSTGRADUATE SCHOOL MONTEREY CA, Tech. Rep., 2006.

\bibitem{gate-redundancy-2009}
A.~Namazi and M.~Nourani, ``Gate-level redundancy: A new design-for-reliability
  paradigm for nanotechnologies,'' \emph{IEEE transactions on very large scale
  integration (VLSI) systems}, vol.~18, no.~5, pp. 775--786, 2009.

\bibitem{TMR-1962}
R.~E. Lyons and W.~Vanderkulk, ``The use of triple-modular redundancy to
  improve computer reliability,'' \emph{IBM journal of research and
  development}, vol.~6, no.~2, pp. 200--209, 1962.

\bibitem{TMR-automated-2018}
L.~A.~C. Benites and F.~L. Kastensmidt, ``Automated design flow for applying
  triple modular redundancy (tmr) in complex digital circuits,'' in \emph{2018
  IEEE 19th Latin-American Test Symposium (LATS)}.\hskip 1em plus 0.5em minus
  0.4em\relax IEEE, 2018, pp. 1--4.

\bibitem{TMR-FPGA-comparison2007}
K.~S. Morgan, D.~L. McMurtrey, B.~H. Pratt, and M.~J. Wirthlin, ``A comparison
  of tmr with alternative fault-tolerant design techniques for fpgas,''
  \emph{IEEE transactions on nuclear science}, vol.~54, no.~6, pp. 2065--2072,
  2007.

\bibitem{gate-res-2016}
X.~Han, M.~Donato, R.~I. Bahar, A.~Zaslavsky, and W.~Patterson, ``Design of
  error-resilient logic gates with reinforcement using implications,'' in
  \emph{Proceedings of the 26th edition on Great Lakes Symposium on VLSI},
  2016, pp. 191--196.

\bibitem{evolution-circuit-evaluation-1999}
J.~D. Lohn and S.~P. Colombano, ``A circuit representation technique for
  automated circuit design,'' \emph{IEEE Transactions on Evolutionary
  Computation}, vol.~3, no.~3, pp. 205--219, 1999.

\bibitem{sinw_array}
D.~Jeon, S.~Park, S.~Pregl, T.~Mikolajick, and W.~Weber, ``Reconfigurable
  thin-film transistors based on a parallel array of si-nanowires,'' vol. 129,
  pp. 1\,245\,041 -- 1\,245\,049, 2021.

\bibitem{3D-book-2017}
V.~F. Pavlidis, I.~Savidis, and E.~G. Friedman, \emph{Three-dimensional
  integrated circuit design}.\hskip 1em plus 0.5em minus 0.4em\relax Newnes,
  2017.

\bibitem{soc2001}
G.~Martin and H.~Chang, ``System-on-chip design,'' in \emph{ASICON 2001. 2001
  4th International Conference on ASIC Proceedings (Cat. No. 01TH8549)}.\hskip
  1em plus 0.5em minus 0.4em\relax IEEE, 2001, pp. 12--17.

\bibitem{mpsoc-2008}
W.~Wolf, A.~A. Jerraya, and G.~Martin, ``Multiprocessor system-on-chip (mpsoc)
  technology,'' \emph{IEEE Transactions on Computer-Aided Design of Integrated
  Circuits and Systems}, vol.~27, no.~10, pp. 1701--1713, 2008.

\bibitem{diverse-virtualization-2008}
B.-G. Chun, P.~Maniatis, and S.~Shenker, ``Diverse replication for
  single-machine byzantine-fault tolerance.'' in \emph{USENIX Annual Technical
  Conference}, 2008, pp. 287--292.

\bibitem{BFT-SDN-2016}
K.~ElDefrawy and T.~Kaczmarek, ``Byzantine fault tolerant software-defined
  networking (sdn) controllers,'' in \emph{2016 IEEE 40th annual computer
  software and applications conference (COMPSAC)}, vol.~2.\hskip 1em plus 0.5em
  minus 0.4em\relax IEEE, 2016, pp. 208--213.

\bibitem{intrusion-virtualization2010}
V.~S. J{\'u}nior, L.~C. Lung, M.~Correia, J.~da~Silva~Fraga, and J.~Lau,
  ``Intrusion tolerant services through virtualization: A shared memory
  approach,'' in \emph{2010 24th IEEE International Conference on Advanced
  Information Networking and Applications}.\hskip 1em plus 0.5em minus
  0.4em\relax IEEE, 2010, pp. 768--774.

\bibitem{gpgpu-CUDA-2012gpgpu}
J.~Ghorpade, J.~Parande, M.~Kulkarni, and A.~Bawaskar, ``Gpgpu processing in
  cuda architecture,'' \emph{arXiv preprint arXiv:1202.4347}, 2012.

\bibitem{vPLC-2021}
\BIBentryALTinterwordspacing
``Virtualized programmable logic controllers,'' 2021, accessed on: Feb, 14,
  2023. [Online]. Available:
  \url{controleng.com/articles/virtualized-programmable-logic-controllers/}
\BIBentrySTDinterwordspacing

\bibitem{plc-sec2021security}
J.~Hajda, R.~Jakuszewski, and S.~Ogonowski, ``Security challenges in industry
  4.0 plc systems,'' \emph{Applied Sciences}, vol.~11, no.~21, p. 9785, 2021.

\bibitem{ecu-2021survey}
C.~Wulf, M.~Willig, and D.~G{\"o}hringer, ``A survey on hypervisor-based
  virtualization of embedded reconfigurable systems,'' in \emph{2021 31st
  International Conference on Field-Programmable Logic and Applications
  (FPL)}.\hskip 1em plus 0.5em minus 0.4em\relax IEEE, 2021, pp. 249--256.

\bibitem{xilinx-IP-core}
\BIBentryALTinterwordspacing
I.~Advanced Micro~Devices, ``Amd/xilinx intellectual property,'' 2023, accessed
  on: May 1st, 2023. [Online]. Available:
  \url{https://www.xilinx.com/products/intellectual-property.html}
\BIBentrySTDinterwordspacing

\bibitem{Intel-IP-core}
\BIBentryALTinterwordspacing
I.~Corporation, ``Intel fpga intellectual property,'' 2023, accessed on: May
  1st, 2023. [Online]. Available:
  \url{https://www.intel.com/content/www/us/en/products/details/fpga/intellectual-property.html}
\BIBentrySTDinterwordspacing

\bibitem{hdl-1998jhdl}
P.~Bellows and B.~Hutchings, ``Jhdl-an hdl for reconfigurable systems,'' in
  \emph{Proceedings. IEEE symposium on FPGAs for custom computing machines
  (Cat. No. 98TB100251)}.\hskip 1em plus 0.5em minus 0.4em\relax IEEE, 1998,
  pp. 175--184.

\bibitem{passive-backup-1993primary}
N.~Budhiraja, K.~Marzullo, F.~B. Schneider, and S.~Toueg, ``The primary-backup
  approach,'' \emph{Distributed systems}, vol.~2, pp. 199--216, 1993.

\bibitem{semi-passive-rep-1998}
X.~Defago, A.~Schiper, and N.~Sergent, ``Semi-passive replication,'' in
  \emph{Proceedings Seventeenth IEEE Symposium on Reliable Distributed Systems
  (Cat. No.98CB36281)}, 1998, pp. 43--50.

\bibitem{1253179}
D.~Ernst, N.~S. Kim, S.~Das, S.~Pant, R.~Rao, T.~Pham, C.~Ziesler, D.~Blaauw,
  T.~Austin, K.~Flautner, and T.~Mudge, ``Razor: a low-power pipeline based on
  circuit-level timing speculation,'' in \emph{Proceedings. 36th Annual
  IEEE/ACM International Symposium on Microarchitecture, 2003. MICRO-36.},
  2003, pp. 7--18.

\bibitem{1610623}
S.~Das, D.~Roberts, S.~Lee, S.~Pant, D.~Blaauw, T.~Austin, K.~Flautner, and
  T.~Mudge, ``A self-tuning dvs processor using delay-error detection and
  correction,'' \emph{IEEE Journal of Solid-State Circuits}, vol.~41, no.~4,
  pp. 792--804, 2006.

\bibitem{6487728}
S.~Kim, I.~Kwon, D.~Fick, M.~Kim, Y.-P. Chen, and D.~Sylvester, ``Razor-lite: A
  side-channel error-detection register for timing-margin recovery in 45nm soi
  cmos,'' in \emph{2013 IEEE International Solid-State Circuits Conference
  Digest of Technical Papers}, 2013, pp. 264--265.

\bibitem{smr:1993}
F.~B. Schneider, \emph{Replication Management Using the State-Machine
  Approach}.\hskip 1em plus 0.5em minus 0.4em\relax USA: ACM
  Press/Addison-Wesley Publishing Co., 1993, p. 169–197.

\bibitem{lamport2001paxos}
L.~Lamport, ``Paxos made simple,'' \emph{ACM SIGACT News (Distributed Computing
  Column) 32, 4 (Whole Number 121, December 2001)}, pp. 51--58, 2001.

\bibitem{efficient-hybrid-BFT-2016}
T.~Distler, C.~Cachin, and R.~Kapitza, ``Resource-efficient byzantine fault
  tolerance,'' \emph{IEEE Transactions on Computers}, vol.~65, no.~9, pp.
  2807--2819, 2016.

\bibitem{minBFT-2011efficient}
G.~S. Veronese, M.~Correia, A.~N. Bessani, L.~C. Lung, and P.~Verissimo,
  ``Efficient byzantine fault-tolerance,'' \emph{IEEE Transactions on
  Computers}, vol.~62, no.~1, pp. 16--30, 2011.

\bibitem{pistis:2021}
D.~Kozhaya, J.~Decouchant, V.~Rahli, and P.~Esteves-Verissimo, ``Pistis: an
  event-triggered real-time byzantine-resilient protocol suite,'' \emph{IEEE
  Transactions on Parallel and Distributed Systems}, vol.~32, no.~9, pp.
  2277--2290, 2021.

\bibitem{IRS-vehicle-2022intrusion}
A.~Shoker, V.~Rahli, J.~Decouchant, and P.~Esteves-Verissimo, ``Intrusion
  resilience systems for modern vehicles,'' in \emph{In the 97th IEEE Vehicular
  Technology Conference (VTC2023)}.\hskip 1em plus 0.5em minus 0.4em\relax
  IEEE, 2023.

\bibitem{autosar-2023}
\BIBentryALTinterwordspacing
``Autosar standard,'' 2023, accessed on: Feb, 14, 2023. [Online]. Available:
  \url{https://www.autosar.org/}
\BIBentrySTDinterwordspacing

\bibitem{opengl}
\BIBentryALTinterwordspacing
K.~Group, ``Opengl,'' 2023, accessed on: April, 19, 2023. [Online]. Available:
  \url{https://www.opengl.org/}
\BIBentrySTDinterwordspacing

\bibitem{risc-v}
\BIBentryALTinterwordspacing
R.-V. International, ``Risc-v,'' 2023, accessed on: April, 19, 2023. [Online].
  Available: \url{https://riscv.org/}
\BIBentrySTDinterwordspacing

\bibitem{5479525}
J.~Zeppenfeld, A.~Bouajila, A.~Herkersdorf, and W.~Stechele, ``Towards
  scalability and reliability of autonomic systems on chip,'' in \emph{2010
  13th IEEE International Symposium on Object/Component/Service-Oriented
  Real-Time Distributed Computing Workshops}, 2010, pp. 73--80.

\bibitem{soft-rejuvination-1995software}
Y.~Huang, C.~Kintala, N.~Kolettis, and N.~D. Fulton, ``Software rejuvenation:
  Analysis, module and applications,'' in \emph{Twenty-fifth international
  symposium on fault-tolerant computing. Digest of papers}.\hskip 1em plus
  0.5em minus 0.4em\relax IEEE, 1995, pp. 381--390.

\bibitem{split-2020secure}
Y.~Yang, Z.~Chen, Y.~Liu, T.-Y. Ho, Y.~Jin, and P.~Zhou, ``How secure is split
  manufacturing in preventing hardware trojan?'' \emph{ACM Transactions on
  Design Automation of Electronic Systems (TODAES)}, vol.~25, no.~2, pp. 1--23,
  2020.

\bibitem{soverign-2022digital}
A.~Shoker, ``Digital sovereignty strategies for every nation,'' 2022.

\bibitem{Soc-rejuvination-2023system}
A.~T. Sheikh, A.~Shoker, and P.~Esteves-Verissimo, ``System on chip
  rejuvenation in the wake of persistent attacks,'' in \emph{the 16th European
  Workshop on Systems Security (EuroSec), EuroSys-W}.\hskip 1em plus 0.5em
  minus 0.4em\relax IEEE, 2023.

\bibitem{adaptive-ft-smr2021}
\BIBentryALTinterwordspacing
D.~Silva, R.~Graczyk, J.~Decouchant, M.~Volp, and P.~Esteves-Verissimo,
  ``Threat adaptive byzantine fault tolerant state-machine replication,'' in
  \emph{2021 40th International Symposium on Reliable Distributed Systems
  (SRDS)}.\hskip 1em plus 0.5em minus 0.4em\relax Los Alamitos, CA, USA: IEEE
  Computer Society, sep 2021, pp. 78--87. [Online]. Available:
  \url{https://doi.ieeecomputersociety.org/10.1109/SRDS53918.2021.00017}
\BIBentrySTDinterwordspacing

\bibitem{adapt-2015making}
J.-P. Bahsoun, R.~Guerraoui, and A.~Shoker, ``Making bft protocols really
  adaptive,'' in \emph{2015 IEEE International Parallel and Distributed
  Processing Symposium}.\hskip 1em plus 0.5em minus 0.4em\relax IEEE, 2015, pp.
  904--913.

\bibitem{adapt-SDN-2018morph}
E.~Sakic, N.~{\DH}eri{\'c}, and W.~Kellerer, ``Morph: An adaptive framework for
  efficient and byzantine fault-tolerant sdn control plane,'' \emph{IEEE
  Journal on Selected Areas in Communications}, vol.~36, no.~10, pp.
  2158--2174, 2018.

\bibitem{gouveia2022behind}
I.~P. Gouveia, M.~V{\"o}lp, and P.~Esteves-Verissimo, ``Behind the last line of
  defense: Surviving soc faults and intrusions,'' \emph{Computers \& Security},
  vol. 123, p. 102920, 2022.

\bibitem{behl2017-hybrid}
J.~Behl, T.~Distler, and R.~Kapitza, ``Hybrids on steroids: Sgx-based high
  performance bft,'' in \emph{Proceedings of the Twelfth European Conference on
  Computer Systems}, 2017, pp. 222--237.

\bibitem{A2M-chun2007attested-hybrid}
B.-G. Chun, P.~Maniatis, S.~Shenker, and J.~Kubiatowicz, ``Attested append-only
  memory: Making adversaries stick to their word,'' \emph{ACM SIGOPS Operating
  Systems Review}, vol.~41, no.~6, pp. 189--204, 2007.

\bibitem{gupta2022dissecting-hybrid}
S.~Gupta, S.~Rahnama, S.~Pandey, N.~Crooks, and M.~Sadoghi, ``Dissecting bft
  consensus: In trusted components we trust!'' \emph{arXiv preprint
  arXiv:2202.01354}, 2022.

\bibitem{kapitza2012cheapbft-hybrid}
R.~Kapitza, J.~Behl, C.~Cachin, T.~Distler, S.~Kuhnle, S.~V. Mohammadi,
  W.~Schr{\"o}der-Preikschat, and K.~Stengel, ``Cheapbft: Resource-efficient
  byzantine fault tolerance,'' in \emph{Proceedings of the 7th ACM european
  conference on Computer Systems}, 2012, pp. 295--308.

\bibitem{teechain-hybrid}
J.~Lind, O.~Naor, I.~Eyal, F.~Kelbert, E.~G. Sirer, and P.~Pietzuch,
  ``Teechain: a secure payment network with asynchronous blockchain access,''
  in \emph{Proceedings of the 27th ACM Symposium on Operating Systems
  Principles}, 2019, pp. 63--79.

\bibitem{jacobtrinc-hybrid}
D.~L. J. R.~D. Jacob and R.~L.~T. Moscibroda, ``Trinc: Small trusted hardware
  for large distributed systems.''

\bibitem{aguilera2023ubft-hybrid}
M.~K. Aguilera, N.~Ben-David, R.~Guerraoui, A.~Murat, A.~Xygkis, and
  I.~Zablotchi, ``ubft: Microsecond-scale bft using disaggregated memory,'' in
  \emph{Proceedings of the 28th ACM International Conference on Architectural
  Support for Programming Languages and Operating Systems, Volume 2}, 2023, pp.
  862--877.

\bibitem{7307998}
T.~Distler, C.~Cachin, and R.~Kapitza, ``Resource-efficient byzantine fault
  tolerance,'' \emph{IEEE Transactions on Computers}, vol.~65, no.~9, pp.
  2807--2819, 2016.

\bibitem{1353018}
M.~Correia, N.~Neves, and P.~Verissimo, ``How to tolerate half less one
  byzantine nodes in practical distributed systems,'' in \emph{Proceedings of
  the 23rd IEEE International Symposium on Reliable Distributed Systems,
  2004.}, 2004, pp. 174--183.

\bibitem{10.1145/3492321.3519568}
\BIBentryALTinterwordspacing
J.~Decouchant, D.~Kozhaya, V.~Rahli, and J.~Yu, ``Damysus: Streamlined bft
  consensus leveraging trusted components,'' in \emph{Proceedings of the
  Seventeenth European Conference on Computer Systems}, ser. EuroSys '22.\hskip
  1em plus 0.5em minus 0.4em\relax New York, NY, USA: Association for Computing
  Machinery, 2022, p. 1–16. [Online]. Available:
  \url{https://doi.org/10.1145/3492321.3519568}
\BIBentrySTDinterwordspacing

\end{thebibliography}

%\newpage

%\input{intro.tex}
%\input{models.tex}
\end{document}